\documentclass[a4paper]{amsart}
\usepackage{amsfonts, amsmath,amsthm,amssymb}
\usepackage{amscd,verbatim}
\usepackage{graphicx}
\usepackage{amsmath}
\usepackage{amssymb}
\usepackage[arrow,matrix,curve]{xy}
\usepackage{graphicx}

\def\alg{{\mathcal A}}

\def\hil{{\mathcal H}}

\newcommand\ZZ{\mathbb{Z}}

\newcommand\bbR{\mathbb{R}}
\newcommand\bbZ{\mathbb{Z}}

\newcommand\bbC{\mathbb{C}}
\newcommand\Sign{{\rm Sign}}

\newcommand{\spec}{\operatorname{spec}}

\newcommand{\End}{\operatorname{End}}
\newcommand{\tr}{\operatorname{tr}}
\newcommand{\Tr}{\operatorname{Tr}}

\newcommand{\cA}{{\mathcal A}}

\newcommand{\cF}{{\mathcal F}}

\newcommand{\cB}{{\mathcal B}}
\newcommand{\cD}{{\mathcal D}}
\newcommand{\cC}{{\mathcal C}}
\newcommand{\cL}{{\mathcal L}}
\newcommand{\cI}{{\mathcal I}}

\newcommand{\cG}{{\mathcal G}}
\newcommand{\cK}{{\mathcal K}}

\newcommand{\cH}{{\mathcal H}}
\newcommand{\cE}{{\mathcal E}}

\newcommand{\cS}{{\mathcal S}}

\theoremstyle{plain}
\newtheorem{theorem}{Theorem}[section]
\newtheorem{theorem*}{Theorem}
\newtheorem{lemma}[theorem]{Lemma}
\newtheorem{proposition}[theorem]{Proposition}
\newtheorem{cor}[theorem]{Corollary}
\newtheorem{cor*}{Corollary}

\theoremstyle{definition}
\newtheorem{definition}[theorem]{Definition}
\newtheorem{definition*}{Definition}

\theoremstyle{remark}

\newtheorem{remarks}[theorem]{Remarks}
\newtheorem{remarks*}{Remarks}

\renewcommand{\subjclassname}{%
\small{\it \textup{2000} Mathematics Subject Classification:}}
\begin{document}
\title[The geometry of determinant line bundles in ncg]
{The geometry of determinant line bundles\\ in noncommutative geometry}

\author[Partha Sarathi Chakraborty]{Partha Sarathi Chakraborty}
\address{Department of Mathematics, University of Adelaide,
Adelaide 5005, Australia\\
(\tiny{On leave from The Institute of Mathematical Sciences, Chennai, India})} 
\email{partha.chakraborty@adelaide.edu.au, parthac@imsc.res.in}

\author[Varghese Mathai]{Varghese Mathai}
\address{Department of Mathematics, University of Adelaide,
Adelaide 5005, Australia}
\email{mathai.varghese@adelaide.edu.au}

\begin{abstract}
This paper is concerned with the study of the geometry of determinant line bundles
associated to families of spectral triples parametrized by the moduli space of 
gauge equivalent classes of Hermitian connections on a Hermitian finite projective module.
We illustrate our results with some examples that arise in noncommutative geometry.

\end{abstract}
\maketitle
\noindent\subjclassname{\,\,\small{58B34 (primary), 46L87, 58G26 (secondary)}\\}

\section*{Introduction}
In the mid 1990s, Connes and Moscovici \cite{CM95} formulated and proved a far reaching local index theorem 
for spectral triples and introduced the correct definition of dimension in the noncommutative setting, 
where it is no longer an integer, but rather a subset of $\bbC$  which is called the dimension spectrum. This paper aims to understand the stability of the dimension spectrum for families of spectral triples,  and its implications on the existence of geometric structures on the 
determinant line bundle,
such as the Quillen metric and the determinant section.

That is,  we are concerned with the study of the geometry of determinant line bundles
associated to families of spectral triples $(\cA, \cH, D)$
parametrized by the moduli space of 
gauge equivalent classes of Hermitian connections on a Hermitian finite projective module.
Recall that spectral triples $(\cA, \cH, D)$ were introduced by Connes \cite{Connes94},
as defining a noncommutative manifolds,
where $\cA$ is a separable, unital $C^*$-algebra acting on a separable
Hilbert space $\hil$, $D$ an unbounded self-adjoint operator on
$\hil$ such that $D$ has compact resolvent, and $[D,a]$ is a bounded operator on
 $\cH$ for all $a\in\alg$.

Given a spectral triple $(\cA, \cH, D)$, a finite projective module $E$ over $\cA$, a Hermitian structure on $E$ and any Hermitian connection $\nabla$ on $E$,
  a basic stability result implies that $(\End_\cA(E), E\otimes_\cA\cH, D_{E, \nabla})$ is again 
 a spectral triple. 
 The space of all Hermitian connections on $E$ is an affine space 
 $\cC_E$, and the gauge group 
 $\cG$ is defined to be a Lie subgroup of the group 
${\rm Aut}_\cA(E)$ of invertible elements in $\End_\cA(E)$,
such that $\cG$ acts smoothly and freely on  $\cC_E$. 
 Analogous to the classical case we define the determinant line bundle $\cL$ of the index bundle for this family 
of spectral triples. This is a line bundle on 
the moduli space $\cC_E/\cG$ of gauge equivalent classes of 
Hermitian connections on $E$. 

In order to state the hypotheses required to define the Quillen metric \cite{Quillen85} and the determinant section 
of $\cL$, we need to utilise the notions of  regularity and {simple dimension spectrum} 
introduced in \cite{CM95}. 
More precisely, we require the spectral triple 
$(\cA, \cH, D)$ to be { regular}  with simple  { dimension spectrum} and zero
is not in the { dimension spectrum}. These are precisely the same assumptions 
made by Connes and Chamseddine \cite{CC06} in their work on inner fluctuations of spectral actions,
except that we do not need the assumption that zero is not in the spectrum of $D$, cf. \S\ref{sect:zero}. 
In this context, Higson \cite{Hi00} also treats the case when zero is in the spectrum of $D$, but our approach
differs from his.
Another technical result proved here is the stability 
property for regular spectral triples $(\cA, \cH, D)$, which says that
$(\End_\cA(E), E\otimes_\cA\cH, D_{E, \nabla})$ is again a regular spectral triple with simple dimension spectrum
for any Hermitian structure on $E$ and any Hermitian connection $\nabla$ on $E$.
For the application of these constructions to a mathematical understanding of anomalies,
that is, the nonpreservation of a symmetry of the classical action by the full quantum action in a
gauge field theory, see \cite{AS84,Freed86}.

The last section works out an explicit calculation of the Quillen metric and determinant section of the  
determinant line bundle of the index bundle for the family of spectral triples on the noncommutative torus
parametrized by the moduli space of flat Yang-Mills connections on a free module of rank one
which were studied by Connes-Rieffel \cite{CR}. These are expressed in terms of the theta and eta
functions on the moduli space which is a torus.

In \cite{Perrot99},  Perrot has studied a K-theoretic index theorem 
for families of spectral triples parametrized by the moduli space of 
gauge equivalent classes of Hermitian connections on $E$. He makes the 
restrictive assumption that the gauge group $\cG$ be contained in the unitary group of $\cA$, 
which is unnecessary in our context here, and he also does not construct the determinant line bundle of the 
index bundle and its geometry, which is the main study in this paper. 

\section{Preliminaries}
In this section, we recall the construction of the determinant line on the Banach manifold of 
all bounded Fredholm operators acting between separable Hilbert spaces. This motivates the 
constructions used later in the paper.

Recall that a Fredholm operator 
$T : \cH_0 \longrightarrow \cH_1$
between two infinite dimensional Hilbert spaces $\cH_0, \cH_1$, is a bounded linear operator
such that $\dim({\rm ker}(T)) < \infty$ and $ \dim({\rm coker}(T))< \infty$. This implies in particular
that ${\rm Im}(T)$ is a closed subspace of $\cH_1$.
Let $\cF = \cF(\cH_0, \cH_1)$ denote the space of all Fredholm operators between two infinite dimensional Hilbert spaces $\cH_0, \cH_1$. 
It follows from Atkinson's theorem that $\cF$ is an open subset of the Banach space $\cB(\cH_0, \cH_1)$,
of all bounded linear operators between the two Hilbert spaces,
which establishes in particular that $\cF$
is a Banach manifold modeled on the Banach space $\cB(\cH_0, \cH_1)$.
It has countably many connected
components labelled by, ${\rm index} : \pi_0(\cF) \stackrel{\cong}{\to} \bbZ$, where for $T\in \cF$,
$${\rm index}(T) = \dim({\rm ker}(T)) -  \dim({\rm coker}(T)).$$ 
We want to review the construction of a smooth line bundle
${\rm DET} \to~ \cF,$ called the {\em determinant line bundle}, such that ${\rm DET}_T = \Lambda^{max}({\rm ker}(T))^* \otimes  \Lambda^{max}({\rm coker}(T))$.
The obvious definition does not work since $\dim({\rm ker}(T)) $ jumps as $T$ varies smoothly.

This problem was essentially fixed in \cite{Quillen85}. For the convenience of the reader we elaborate on his solution.   Let 
$ {\rm Gr}_{\rm fin}$ denote the space of all finite dimensional subspaces of $\cH_1$.
Consider the open cover $\{U_F\}$ of $\cF$, where $F\in  {\rm Gr}_{\rm fin}$ and
$U_F = \{ T\in \cF: {\rm Im}(T) + F = \cH_1\}$. For $T\in U_F$, consider the exact
sequence of finite dimensional vector spaces,
\begin{equation}\label{eqn:index1}
0\to {\rm ker}(T) \to T^{-1}F \stackrel{T}{\to} F \to {\rm coker}(T)\to 0.
\end{equation}
Since index is constant on smooth families, and the rank of $F$ is fixed on $U_F$,
therefore the rank of
$T^{-1}F$ is constant on smooth families in $U_F$.
So $\cE^F\to U_F$
defined by $\cE^F_T = T^{-1}F$, is a smooth vector bundle.
The virtual vector bundle ${\rm INDEX}^F \to U_F$ is defined to be the
pair $(\cE^F, F)$, where $F$ denotes the trivial vector bundle
over $U_F$ with fibre $F$.

Using the inner products on $\cH_i, \, i=0,1$, the sequence in equation 
\eqref{eqn:index1} splits,
$ {\rm ker}(T) \oplus F \cong T^{-1}F \oplus {\rm coker}(T)$, therefore
$$
\Lambda^{max}({\rm ker}(T))^* \otimes  \Lambda^{max}({\rm coker}(T))\cong
\Lambda^{max} (T^{-1}F)^* \otimes   \Lambda^{max} F.
$$

The {\em determinant line
bundle}, ${\rm DET}^F \to U_F$, is defined as the smooth line bundle, 
${\rm DET}^F = {\rm det}(({\rm INDEX}^F)^*)$, i.e. 
${\rm DET}^F = \Lambda^{max} (\cE^F)^*
\otimes  \Lambda^{max} F$.

Suppose that $T\in U_E\cap U_F$. Then we have the exact sequences

\begin{eqnarray*}
\xymatrix{
0 \ar[r]&   {\rm ker}(T)   \ar[d]_{=} \ar^-{\iota_1}[r] & T^{-1}F \oplus (E\setminus F)\ar[d]_{\varphi} \ar[r]^-{T\oplus1} &  F +E \ar[d]_{=} \ar[r]&  {\rm coker}(T)  \ar[d]_{=}  \ar[r]& 0\\
0 \ar[r]&  {\rm ker}(T)  \ar[r]^-{\iota_2} & T^{-1}E \oplus (F\setminus E) \ar[r]^-{T\oplus 1} & F +E \ar[r] &  {\rm coker}(T)  \ar[r]& 0,
   }
\end{eqnarray*}
where $\varphi$ is uniquely defined so as to make the diagram commute. Here by $ (E\setminus F)$
we mean the quotient, $E/(E\cap F)$ etc..
By the Five Lemma, the map $\varphi$ is an isomorphism, and 
therefore, ${\rm DET}^F = \Lambda^{max} (\cE^F)^*\otimes  \Lambda^{max}  (E\setminus F)^*
\otimes  \Lambda^{max} (F+E)$ and ${\rm DET}^E = 
\Lambda^{max} (\cE^E)^*\otimes  \Lambda^{max}  (F\setminus E)^*
\otimes  \Lambda^{max} (F+E)$ are naturally isomorphic via $\Lambda^{max}\varphi\otimes 1$ 
over $U_E\cap U_F$. Here we have used the fact that $ \Lambda^{max}  (V)^*
\otimes  \Lambda^{max}  (V)$ is canonically trivial for any vector bundle $V$.
By the clutching construction, it follows that 
 ${\rm DET}$ defines a smooth line bundle over the union $U_E\cup U_F$.

To show that ${\rm DET}$ defines a smooth line bundle over the whole of $\cF$, 
we need to show that on triple overlaps $U_E \cap U_F \cap U_G$, there are natural
isomorphisms between the determinant line bundles ${\rm DET}^E$, ${\rm DET}^F $ and ${\rm DET}^G$.
The proof is similar to the case of double overlaps.
Suppose that $T\in U_E\cap U_F \cap U_G$. Then we have the exact sequences

\begin{eqnarray*}
\xymatrix{
0 \ar[r]&   {\rm ker}(T)   \ar[d]_{=} \ar^-{\iota_1}[r] & T^{-1}G \oplus (E +F\setminus G)\ar[d]_{\varphi_1} \ar[r]^-{T\oplus1} &  E +F +G \ar[d]_{=} \ar[r]&  {\rm coker}(T)  \ar[d]_{=}  \ar[r]& 0\\
0 \ar[r]&   {\rm ker}(T)   \ar[d]_{=} \ar^-{\iota_1}[r] & T^{-1}F \oplus (E+G\setminus F)\ar[d]_{\varphi_2} \ar[r]^-{T\oplus1} &  E +F +G \ar[d]_{=} \ar[r]&  {\rm coker}(T)  \ar[d]_{=}  \ar[r]& 0\\
0 \ar[r]&  {\rm ker}(T)  \ar[r]^-{\iota_2} & T^{-1}E \oplus (F+G\setminus E) \ar[r]^-{T\oplus 1} & E +F+G \ar[r] &  {\rm coker}(T)  \ar[r]& 0,
   }
\end{eqnarray*}
where $\varphi_j, \, j=1,2,$ is uniquely defined so as to make the diagram commute.
By the Five Lemma, the map $\varphi_j, \, j=1,2,$ is an isomorphism, and 
therefore, $${\rm DET}^G = \Lambda^{max} (\cE^G)^*\otimes  \Lambda^{max}  (E+F\setminus G)^*
\otimes  \Lambda^{max} (E+F+G)$$ and 
$${\rm DET}^F = \Lambda^{max} (\cE^F)^*\otimes  \Lambda^{max}  (E+G\setminus F)^*
\otimes  \Lambda^{max} (E+F+G)$$ are naturally isomorphic via $\Lambda^{max}\varphi_1\otimes 1$ 
over $U_E\cap U_F \cap U_G$, where we have used the fact that $ \Lambda^{max}  (V)^*
\otimes  \Lambda^{max}  (V)$ is canonically trivial for any vector bundle $V$.
Similarly,  $${\rm DET}^F = \Lambda^{max} (\cE^F)^*\otimes  \Lambda^{max}  (E+G\setminus F)^*
\otimes  \Lambda^{max} (E+F+G)$$ and $${\rm DET}^E = 
\Lambda^{max} (\cE^E)^*\otimes  \Lambda^{max}  (F+G\setminus E)^*
\otimes  \Lambda^{max} (E+F+G)$$ are naturally isomorphic via $\Lambda^{max}\varphi_2\otimes 1$ 
over $U_E\cap U_F\cap U_G$. Therefore ${\rm DET}^G, \, {\rm DET}^F$ and ${\rm DET}^E$
are canonically identified on the overlaps $U_E\cap U_F\cap U_G$.
By the clutching construction, it follows that  ${\rm DET}$ defines a consistent smooth line bundle 
over the union $U_E\cup U_F \cup U_G$. Since $E, F$ and $G$ are arbitrary finite dimensional subspaces 
of $\cH_1$, it follows that ${\rm DET}$ defines a smooth line bundle
on the Banach manifold $\cF$.

It is a fact that $\cF$ is homotopy equivalent to $\bbZ \times BU(\infty)$, so that
by Bott periodicity, $\pi_{2j}(\cF)\cong \bbZ$ and  $\pi_{2j+1}(\cF) = \{0\}$ for $j\ge 0$.
So by Hurewicz's theorem, $H^2(\cF_0, \bbZ) \cong \bbZ$, where $\cF_0$ is the
connected component of $\cF$ consisting of Fredholm operators of index equal to zero. 
It is a fact that the first Chern class,
$c_1({\rm DET})$ is the generator of $H^2(\cF_0, \bbZ) \cong \bbZ$.

\section{Stability of spectral triples coupled to a Hermitian finite projective module with Hermitian connection}

The notion of a spectral triple was introduced by
Connes~\cite{Connes94} as the criteria defining a noncommutative
spin geometry. A slightly weakened version, dropping the postulate of
the existence of a real structure, was given by
Moscovici~\cite{Moscovici97} which was suited to describe more general 
Poincar\'e dual pairs of algebras, and in particular noncommutative spin$^c$
geometries. 

Let $A$ be a separable, unital $C^*$-algebra acting on a separable
Hilbert space $\hil$. Let $D$ be an unbounded self-adjoint operator on
$\hil$. Let $\alg$ be a smooth unital subalgebra of $A$.
Then $(\alg,\hil,D)$ is said to be a \emph{spectral
 triple} if $D$ has compact resolvent, and $[D,a]$ is a bounded operator on
 $\cH$ for all $a\in\alg$.
$(\cA, \cH, D)$ is said to be {\em even} if there is a self-adjoint involution on $\cH$ with
 respect to which the action of $\alg$ is even and $D$ is an odd operator.
 Otherwise the spectral triple is said to be {\em odd}.
We begin by considering the even case. 


Let $E$ be a finite projective (right-)module 
over $\cA$.  
A {\em Hermitian structure} on 
$E$ is a sesquilinear map 
$( , ) : E \times E \longrightarrow A $
satisfying the following conditions: 
\begin{enumerate}
\item $(\xi a, \eta b) = a^* (\xi , \eta) b, 
\forall \xi , \eta \in E$ and $\forall  a, b \in A$; 
\item $(\xi , \xi ) \ge 0$; 
\item $E$ is self-dual with respect to $( , )$. 
\end{enumerate}

Consider the special case of the free $\cA$-module $E_0 = \cA^q$
which has a canonical Hermitian structure given by 
$(\xi , \eta) = \sum^q_1 \xi_j^* \eta_j ,\qquad 
\forall \xi = (\xi_1, \ldots, \xi_q) , \eta = (\eta_1, \ldots, \eta_q) \in E_0$.

Let $E$ be a finite projective (right) module 
over $\cA$. If we write $E$ as a direct summand $E = e \cA^q$ 
of a free module $E_0$, where $e \in M_q(\cA)$ is a self
adjoint projection.
Then $E$ has a Hermitian structure which is obtained by 
restricting the Hermitian structure on $E_0$ defined above, 
to $E$. That is, every finite projective module $E$ over $\cA$
has a Hermitian structure.

Consider the $\cA$-bimodule of 1-forms on $\cA$, 
$\Omega_D^1(\cA) := \left\{\sum a_j [D, b_j ] ; a_j , b_j \in  \cA \right\}.$
A {\em Hermitian connection} on 
$E$ is a $\bbC$-linear map $\nabla : E \to E \otimes_\cA \Omega_D^1(\cA)$
satisfying the following:
\begin{enumerate}
\item $\nabla(\xi a) = \nabla(\xi ) a + \xi \otimes da, \, \forall \xi \in E , a \in \cA$, (Leibnitz property)
\item $(\xi ,\nabla \eta)  - (\nabla \xi, \eta)=d (\xi, \eta), \, \forall \xi , \eta \in E$; where $da = [D, a], \, \forall a \in \cA$, 
(Hermitian property)
\end{enumerate}
If $ \nabla(\xi ) = \sum\xi_j\otimes \omega_j $, with $\xi_j \in E, \omega_j \in \Omega_D^1(\cA)$,
then $ (\nabla \xi, \eta)=\sum \omega_j^* (\xi_j , \eta)$.

An example of a Hermitian connection is the Grassmannian connection 
$\nabla_0$ on $E = e\cA^q$ given by 
$\nabla_0(\xi) = e d\xi$, where $d\xi= (d \xi_1, \ldots d\xi_q)$. 
Also, any two Hermitian connections differ by a selfadjoint element of 
$\End_\cA(E) \otimes_\cA \Omega_D^1(\cA)$. That is, the space of all
Hermitian connections on $E$ is an affine space $\cC_E$ with associated vector
space $\End_\cA(E) \otimes_\cA \Omega_D^1(\cA)$.

Given a Hermitian finite projective module 
$E$ over $\cA$, we can form the Hilbert space 
$E \otimes_\cA \cH$, by completing the algebraic tensor product with respect to the inner product
$\langle \xi_1 \otimes \eta_1 ,  \xi_2 \otimes \eta_2\rangle = 
\langle \eta_1 , (\xi_1 , \xi_2 ) \eta_2 \rangle \, \forall \xi_j \in E, \eta_j \in \cH$. 

Given such a pair $(E, \nabla)$, one can define a twisted operator 
$D_{E, \nabla}$ on the Hilbert space $E \otimes_\cA \cH$ by setting 
$D_{E, \nabla}(\xi \otimes \eta) = \xi \otimes D(\eta) + \nabla (\xi )\eta, \, \forall \xi \in E , \eta \in \cH.$
Here, if $\nabla (\xi ) = \sum \xi_j \otimes \omega_j$, then 
$\nabla (\xi )\eta =  \sum \xi_j \otimes \omega_j (\eta) \in E \otimes_\cA \cH$.

The we have the following.

\begin{lemma}\label{lemma:compactresolvent}
Let $T$ be an unbounded self-adjoint operator on a Hilbert space $\cH$
and suppose that there is an operator $S \in \cI$, 
where $\cI$ denotes either the ideal of all compact
operators on $\cH$ or a Schatten ideal on $\cH$, 
such that $(iI + T)S - I \in \cI$. Then $(iI+T)^{-1} \in \cI$.
Moreover, $(iI + R+T)S - I \in \cI$ for any bounded self-adjoint operator $R$,
and therefore  $(iI+R+T)^{-1} \in \cI$.

\end{lemma}

\begin{proof}
To prove this, note that 
$$
(iI+T)^{-1} = S - K(iI+T)^{-1} 
$$
where $K = (iI + T)S - I \in \cI$. Therefore the right hand side is in 
the ideal $\cI$ as claimed. Since $RS \in \cI$
for any bounded self-adjoint operator $R$, 
it follows that $(iI + R+T)S - I \in \cI$,
and therefore  $(iI+R+T)^{-1} \in \cI$ as claimed.

\end{proof}

\begin{proposition}[Stability of Spectral Triples I]\label{prop:enhancedST}
Let $(\alg,\hil,D)$ be a {spectral triple}, $E$ be a finite projective $\cA$-module
with Hermitian structure and Hermitian connection $\nabla$. Then 
$(\End_\cA(E),E\otimes_\cA \hil,D_{E, \nabla})$ is also a {spectral triple}. Moreover if  $(\alg,\hil,D)$ is $p$-summable then so is $(\End_\cA(E),E\otimes_\cA \hil,D_{E, \nabla})$.
\end{proposition}


\begin{proof}
First observe that for all natural numbers $N$, $(M_N(\bbC) \otimes\alg, 
\bbC^N \otimes\hil,1\otimes D)$
is a spectral triple, where $ \bbC^N \otimes\hil \cong \alg^N \otimes_\alg \hil$. So the 
result is true when $E$ is a free module and with the trivial connection.
For simplicity of notation, denote by $D$ the operator $I\otimes D$. 
An arbitrary connection on the free module $E$ is of the form 
$D + \sum_i a_i [D, a_i']$, where the term $ \sum_i a_i [D, a_i'], 
\, a_i, a_i' \in \cA,$ is a 
bounded operator which we denote by $R$. Such perturbations of $D$ will be referred as inner fluctuations. The condition that the 
connection is Hermitian implies that $R$ is self-adjoint. By Lemma
\ref{lemma:compactresolvent}, it follows that $(\alg, \cH, D+R)$ 
is again a spectral triple, which is just the statement that spectral
triples are stable under inner fluctuations.

Next we observe there exists a projection $e \in M_N(\alg)$ for some 
natural number $N$ such that $E= e\cA^N$. We want to show that 
$(\End_\cA(E),E\otimes_\cA \hil,D_{E, \nabla_0})$ is a spectral triple, where
$\nabla_0 = e[D, e]$ is the Grassmann connection and we have used the simplified
notation. Consider the compact operator $S= e(iI + D)^{-1} e$. Then
\begin{align}
e(iI+D)eS & = e(iI+D)e(iI + D)^{-1} e\\
& = e[(iI+D), e]e(iI + D)^{-1} e + e
\end{align}
so that $e(iI+D)eS - e \in \cK$. By Lemma
\ref{lemma:compactresolvent}, it follows that 
$e(iI + D)^{-1} e = (ie + eDe)^{-1}$ is a compact operator, 
that is $(\End_\cA(E),E\otimes_\cA \hil,D_{E, \nabla_0})$ is a spectral triple. 
This along with the stability of spectral 
triples under inner fluctuations as proved above implies the result for arbitrary connections on $E$.
Replacing the ideal of compact operators $\cK$ by Schatten ideals we get the summability result.
\end{proof}

\section{Universal families of spectral triples and determinant line bundles}

We follow a setup that is analogous to that of Freed \cite{Freed86}, where it was done
in the classical case of families of Dirac operators on a smooth manifold. 
This noncommutative geometry 
data will enable us to define the determinant line bundle associated to a 
natural family of spectral triples. 
\\

\subsection{Noncommutative Geometry Data for determinant line bundles:}\label{det-assum}
\begin{enumerate}
\item A {\em spectral triple},  $(\alg,\hil,D)$;
\item  A {\em Hermitian finite projective module} $E$ over $\cA$;
\item A  {\em gauge group} $\cG$ which is a Lie subgroup of the group 
${\rm Aut}_\cA(E)$ of automorphisms of $E$,
such that $\cG$ acts smoothly and freely on  $\cC_E$ as follows.
For $\xi \in  E$, $\nabla \in \cC_E$ and $g \in \cG$, define $g.\nabla(\xi) := (g^{-1} \otimes 1)\nabla(g.\xi)$;
\end{enumerate}

\begin{remarks}
\begin{enumerate}
\item There is an associated $\bbZ_2$-graded Hilbertian bundle $\cE \otimes_\cA \cH$ over $\cC_E/\cG$. 
\item 
Therefore one obtains a family 
of spectral triples,  $(\End_\cA(E),E\otimes_\cA \hil, D_{E, \nabla^y})$,  $y \in \cC_E/\cG$. For ease of notation,
denote $D_y = D_{E, \nabla_y}$. Then $D$ can be viewed as an odd degree bundle map $D : \cE \otimes_\cA \cH
\to \cE \otimes_\cA \cH$. 
\item Under our assumptions, the quotient 
$\cC_E/\cG$ is a smooth manifold. If $B$ is a compact smooth finite dimensional submanifold 
of $\cC_E/\cG$, then all of the noncommutative data restricts to $B$.
\item In the case of the spectral triple for a spin manifold, this is just the data for a 
family of twisted Dirac operators.
\end{enumerate}
\end{remarks}


The construction of the determinant line bundle due to Quillen \cite{Quillen85} is 
briefly adapted in our context in the rest of the section. Consider the 
$\cG$-equivariant family of spectral triples 
$\left\{(\End_\alg(E),E\otimes_\cA \hil,D_{E, \nabla}): \, \nabla \in \cC_E\right\}$. 
The $\cG$-equivariant family of 
finite-dimensional spaces $\ker D_{E, \nabla}^+\subset E\otimes_\cA \hil^+$ 
and $\ker D_{E, \nabla}^-\subset E\otimes_\cA \hil^-$ defines a virtual bundle 
${\rm Index}(D_{\cE, \nabla^\cE})$ over $B$, which is the index bundle
for the family, whose fibre at $\nabla$ is
the virtual vector space
$
{\rm Index}(D_{\cE, \nabla^\cE})_\nabla = \ker D_{E, \nabla}^+ -\ker D_{E, \nabla}^-.
$

Now $D_{E, \nabla}^-D_{E, \nabla}^+$ and $D_{E, \nabla}^+D_{E, \nabla}^-$ are self-adjoint operators,
with discrete spectrum, and ${\rm spec}(D_{E, \nabla}^-D_{E, \nabla}^+)=
{\rm spec}(D_{E, \nabla}^+D_{E, \nabla}^-)$.
The simple argument to show this goes as follows.
If $e$ is an eigenvector of
$D_{E, \nabla}^*D_{E, \nabla}$ with eigenvalue
$\lambda$,
then $D_{E, \nabla} e$ is an eigenvector of
$D_{E, \nabla}D_{E, \nabla}^*$ with eigenvalue
$\lambda$.
Also if $f$ is an eigenvector of $D_{E, \nabla}D_{E, \nabla}^*$
with eigenvalue
$\eta$,
then $D_{E, \nabla}^* f$ is an eigenvector of
$D_{E, \nabla}^*D_{E, \nabla}$ with eigenvalue
$\eta$.
This shows that for $\lambda \ne 0$,
$ D_{E, \nabla}$ is an
isomorphism from the $\lambda$-eigenspace of
$D_{E, \nabla}^*D_{E, \nabla}$ to  the $\lambda$-eigenspace of
$D_{E, \nabla} D_{E, \nabla}^*$ (with inverse given by
$\frac{1}{{\lambda}} D_{E, \nabla}^*$).

Let $\; H_{\nabla, \lambda}^+$ denote the span of eigenvectors of 
$D_{E, \nabla}^-D^+_{E, \nabla}$ with eigenvalue $<\lambda$, and 
$\;H_{\nabla, \lambda}^-$ denote the span of eigenvectors of 
$D^+_{E, \nabla} D_{E, \nabla}^-$ with eigenvalue $<\lambda$.
These are smooth vector bundles over the open subset 
$U_\lambda = \{\nabla \in B : \lambda \not\in {\rm spec}(D_{E, \nabla}^-D_{E, \nabla}^+)=
{\rm spec}(D_{E, \nabla}^+D_{E, \nabla}^-)\}$.

It is easy to show that there is an exact sequence,
$$
0 \to \ker(D_{E, \nabla}^+) \to  H_{\nabla, \lambda}^+ \to  H_{\nabla, \lambda}^-
\to  \ker(D_{E, \nabla}^-) \to 0.
$$
This gives rise to a canonical isomorphism,
$$
\Lambda^{max}(\ker(D_{E, \nabla}^+)^*) \otimes\Lambda^{max}  (\ker(D_{E, \nabla}^-))
\cong \Lambda^{max}({H_{\nabla, \lambda}^+}^*) \otimes \Lambda^{max} (H_{\nabla, \lambda}^-)
$$
Therefore we obtain a smooth line bundle $\cL_\lambda:= \Lambda^{max}({H_{\nabla, \lambda}^+}^*) \otimes \Lambda^{max} (H_{\nabla, \lambda}^-)$ over the open set $U_\lambda$. If $\mu>\lambda$, then 
$H_{\nabla, \mu}^\pm = H_{\nabla, \lambda}^\pm \oplus H_{\nabla, \lambda, \mu}^\pm$, where 
$ H_{\nabla, \lambda, \mu}^+$ denotes the span of the eigenvectors of 
$D_{E, \nabla}^-D^+_{E, \nabla}$ with eigenvalues lying in the open interval $(\lambda, \mu)$
and $ H_{\nabla, \lambda, \mu}^-$ denotes the span of the eigenvectors of 
$D_{E, \nabla}^+D^-_{E, \nabla}$ with eigenvalues lying in the open interval $(\lambda, \mu)$. 
Therefore $ \Lambda^{max}H_{\nabla, \mu}^\pm \cong  \Lambda^{max}H_{\nabla, \lambda}^\pm \otimes
\Lambda^{max} H_{\nabla, \lambda, \mu}^\pm$. Since the restriction $D_{E, \nabla}^+\big|_{H_{\nabla, \lambda, \mu}^+}
: H_{\nabla, \lambda, \mu}^+ \to H_{\nabla, \lambda, \mu}^-$ is an isomorphism as observed earlier, 
we deduce that $\det(D_{E, \nabla}^+\big|_{H_{\nabla, \lambda, \mu}^+}):   \Lambda^{max} H_{\nabla, \lambda, \mu}^+ \to 
 \Lambda^{max} H_{\nabla, \lambda, \mu}^-$ is also an isomorphism. Therefore on the overlaps
 $U_\lambda \cap U_\mu$, there is a canonical identification of the determinant line bundles
 $\cL_\lambda$ and $\cL_\mu$ given by $s \mapsto s\otimes \det(D_{E, \nabla}^+\big|_{H_{\nabla, \lambda, \mu}^+})$.
Since $\{U_\lambda: \lambda \in \mathbb Q\}$ is an open cover of $B$, we obtain a global 
determinant line bundle $\cL$ over $B$ associated to the 
$\cG$-equivariant family of spectral triples 
$\left\{(\alg,E\otimes_\cA \hil,D_{E, \nabla}): \, \nabla \in \cC_E\right\}$. 
We state this as a proposition.

\begin{proposition}[Determinant line bundle]\label{prop:det4} 
Let $(\cA, \cH, D)$ be a spectral triple satisfying the 
assumptions in section \ref{det-assum},
and $B$ a smooth compact submanifold of $\cC_E/\cG$.
Then there is a smooth determinant line bundle $\cL$ over $B$ associated to the 
$\cG$-equivariant family of spectral triples 
$\left\{(\End_\alg(E),E\otimes_\cA \hil,D_{E, \nabla}): \, \nabla \in \cC_E\right\}$, 
whose fibre at $\nabla \in B$
is naturally isomorphic to 
$\Lambda^{max}(\ker(D_{E, \nabla}^+)^*) \otimes\Lambda^{max}  (ker(D_{E, \nabla}^-))$.
\end{proposition}

\section{Stability of regular spectral triples, the Quillen metric
and determinant section}

The Quillen metric on the determinant line bundle,
is obtained by patching metrics constructed on the open sets $U_\lambda$. As in the classical case this patching requires a zeta regularization of the metrics on $U_\lambda$. This in turn requires further assumptions on the spectral triple similar to  the ones that 
Connes and Moscovici used. Let $\cH^\infty=\bigcap_{n\ge 1} {\mathfrak Dom}{|D|^n}$; in this section we will assume that every $a \in \cA$ maps $\cH^\infty$ to itself.

\begin{definition} 
\begin{enumerate}
\item 
A spectral triple  $(\cA, \cH, D)$ is said to be
{\em regular} if $\cA$ and $[D,\cA]$ is contained in the domain of $\delta^k$, for all $k$. Here $\delta$ is the derivation $a \mapsto [|D|,a]$. 
\item For a regular spectral triple $(\cA, \cH, D)$
let $\cB_0$ denote the algebra generated by $\gamma, \Sign(D),$ $ \delta^k(a),\delta^k([D, a]), \, a\in \cA, k \ge 0$.
The {\em dimension spectrum} of a $p$-summable regular spectral triple $(\cA, \cH, D)$ 
 is the smallest discrete subset $\Sigma \subset \bbC$ with the
property that all the zeta functions
$$\zeta(s, a) = \Tr(a|D+P|^{-s}), \qquad a \in \cB_0, \quad s\in \bbC, \quad \Re(s) >p $$
have meromorphic continuations to $\bbC$ with poles contained in $ \Sigma$,
where $P$ denotes the orthogonal projection onto the nullspace of $D$.

$(\cA, \cH, D)$ is said to have {\em simple dimension spectrum} if the 
associated zeta functions $\zeta(s, a)$ have only simple poles, for all $a \in \cB_0$.
\end{enumerate}
\end{definition}


Our next stability result is the following.

\begin{proposition}[Stability of Regular Spectral Triples]\label{prop:enhancedRST2}
Let $(\alg,\hil,D)$ be a {regular spectral triple} with simple dimension spectrum 
$\Sigma$, $E$ be a finite projective $\cA$-module
with Hermitian structure and Hermitian connection $\nabla$. 
Then 
$(\End_\cA(E),$ $E\otimes_\cA~\hil,$ $D_{E, \nabla})$ is also a regular spectral triple
with simple dimension spectrum $\Sigma'$ contained in 
$\Sigma -{\mathbb N}$. Moreover if $0 \notin \Sigma$, then $0 \notin \Sigma'$.
\end{proposition}

We defer the proof of this proposition to \S\ref{sect:enhanced} and \S\ref{sect:zero}.

\subsection{Noncommutative Geometry Data for the Quillen metric and 
the determinant section:}\label{det-assum2}
We next modify the noncommutative geometry data given in section \ref{det-assum}, that will enable us to define
the Quillen metric on the determinant line bundle and also define the determinant section.

\begin{enumerate}
\item A {\em regular spectral triple},  $(\alg,\hil,D)$ such that zero is {\em not}
in the simple dimension spectrum;
\item  A {\em Hermitian finite projective module} $E$ over $\cA$;
\item A  {\em gauge group} $\cG$ which is a Lie subgroup of the group 
${\rm Aut}_\cA(E)$ of automorphisms of $E$,
such that $\cG$ acts smoothly and freely on  $\cC_E$ as follows.
For $\xi \in  E$, $\nabla \in \cC_E$ and $g \in \cG$, define $g.\nabla(\xi) := (g^{-1} \otimes 1)\nabla(g.\xi)$;
\end{enumerate}

To simplify notation, we will denote $\zeta(s) := \zeta(s, 1)$. 
Since by the hypotheses above, $0$ is not in the dimension spectrum $\Sigma$ of 
our regular spectral triple $(\cA, \cH, D)$, it enables us to define the derivative at zero $\zeta'(0)$.
In particular, the  regularized determinant $\det(D^*D) = e^{-\zeta'(0)}$ makes sense, 
and is defined to be zero if $0\in \spec(D)$. 
Also, if $\zeta_u(s) =  \Tr(\chi_{[u, \infty]}(|D|)|D'|^{-s})$ for $u>0$ and 
$u\notin \spec(|D|)$ and 
where $\chi_{[u, \infty]}(|D|)$ denotes the spectral projection of $|D|$,
then there is a simple relationship between the two zeta functions,
$$\zeta(s) = \sum_{0<\lambda <u} \frac{1}{\lambda^s} + \zeta_u(s), \quad \lambda \in \spec(|D|)$$
Since $\sum_{0<\lambda <u} \frac{1}{\lambda^s}, \quad \lambda \in \spec(|D|),$ is an entire
function, it follows that $\zeta_u(s)$ also has an analytic continuation to $\bbC\setminus \Sigma$, 
so that in particular, the derivative at zero $\zeta_u'(0)$ is defined.
Moreover, by the stability of regular spectral triples, Proposition \ref{prop:enhancedRST2},
we see that the $\zeta'(0, \nabla)$ and $\zeta_u'(0, \nabla)$ are defined, 
where $\zeta_u(s, \nabla) =  \Tr(\chi_{[u, \infty]}(|D_{E, \nabla}|)|D_{E, \nabla}'|^{-s})$ 
for $u\notin \spec(|D_{E, \nabla}|)$ and 
where $\chi_{[u, \infty]}(|D_{E, \nabla}|)$ denotes the spectral projection of $|D_{E, \nabla}|$
and $\zeta(s, \nabla) = \zeta_0(s, \nabla) $. 


Given the geometric data as given above, we will proceed to first 
construct the Quillen metric on the determinant line bundle $\cL$.
Using the notation of section \ref{det-assum}, we are given a Hermitian inner 
product on $E\otimes_\cA \cH$, which induces Hermitian metrics 
on the vector bundles $\cH_\lambda$, since for each $\nabla \in \cC_E$,
$\cH_{\lambda, \nabla}$ is a finite dimensional subspace of the Hilbert
space $E\otimes_\cA \cH$. Let $g_\mu'$ denote the induced Hermitian
metric on the determinant line bundle $\cL_\lambda$ over the open set
$U_\lambda$. 
Due to the canonical identification of the determinant line bundles
 $\cL_\lambda$ and $\cL_\mu$ over $U_\mu \cap U_\lambda$ given by 
 $s \mapsto s\otimes \det(D_{E, \nabla}^+\big|_{H_{\nabla, \lambda, \mu}^+})$, 
 where $\mu >\lambda$,  we see that 
 $$g_\mu'(s, s) = g_\lambda' (s, s). \left(\prod_{\lambda < \tau< \mu, \, \tau \in \spec(D_E)} \tau\right)^2.$$
Therefore the induced Hermitian metric $g_\lambda'$ does {\em not} define a 
smooth Hermitian metric on $\cL$, but however can be modified as in 
\cite{Quillen85} to give a smooth Hermitian metric $g$  on $\cL$ called
the {\em Quillen metric} as follows. At $\nabla \in U_\lambda$, define 
$g_\lambda = g_\lambda'. e^{-\zeta_\lambda'(0, \nabla)}$. Then for 
$\nabla \in U_\lambda \cap U_\mu$, since $$ e^{-\zeta_\lambda'(0, \nabla)}
=   e^{-\zeta_\mu'(0, \nabla)} \left( \prod_{\lambda < \tau< \mu, \, \tau \in \spec(D_{E, \nabla})} \tau\right),$$
we see that $g_\lambda = g_\mu$ on the overlap $U_\lambda \cap U_\mu$,
showing that $g_\cL := g$ defines a smooth Hermitian metric on the determinant 
line bundle $\cL$.

\begin{proposition}[Quillen metric on the determinant line bundle]\label{prop:det2}
Let $(\cA, \cH, D)$ be a regular spectral triple with dimension spectrum not containing zero.
Let $B$ be a compact submanifold of $\cC_E/\cG$. Then 
there is a smooth Hermitian metric $g_\cL$, called the Quillen metric) 
on the determinant line bundle $\cL$ over $B$ associated to the 
$\cG$-equivariant family of spectral triples 
$\left\{(\End_\alg(E),E\otimes_\cA \hil,D_{E, \nabla}): \, \nabla \in \cC_E\right\}$.
\end{proposition}

We next describe the determinant section of $\cL$ over the open set $U_\lambda \cap U_0$. 
Let $\{\psi_1, \ldots, \psi_n\}$ be a basis of eigenvectors in $\cH_{\lambda}^+$
and  $\{\psi_1^*, \ldots, \psi_n^*\}$ be the dual basis. Then the 
{\em determinant section} is $\det(D^+_{E, \nabla, \lambda}) = 
(\psi_1^*\wedge \ldots \wedge \psi_N^*) \otimes (D\psi_1\wedge \ldots \wedge D\psi_N) $.
On the overlap $U_\lambda \cap U_\mu \cap U_0$, it is easy to see that 
$$\det(D^+_{E, \nabla, \mu}) = \det(D^+_{E, \nabla, \lambda}) \left(\prod_{\lambda < \tau< \mu, \, \tau \in \spec(D_E)} \tau\right).$$
It follows that on the open set $U_0$, there is a smooth section $\det(D_E)$ 
of the determinant line bundle $\cL$. Therefore, 

\begin{proposition}[Determinant section of the determinant line bundle]\label{prop:det3}
Let $(\cA, \cH, D)$ be a regular spectral triple with dimension spectrum not containing zero.
Let $B$ be a compact submanifold of $\cC_E/\cG$. Then 
there is a smooth determinant section $\det(D_E)$ of the determinant line bundle $\cL$,
over $B\cap U_0$ associated to the 
$\cG$-equivariant family of spectral triples 
$\left\{(\End_\alg(E),E\otimes_\cA \hil,D_{E, \nabla}): \, \nabla \in \cC_E\right\}$.
\end{proposition}

 \subsection{Proof of Proposition \ref{prop:enhancedRST2}}\label{sect:enhanced}
 
 We begin by proving the following special case of Proposition \ref {prop:enhancedRST2}.
 
 \begin{proposition}[Stability of Regular Spectral Triples]\label{prop:enhancedRST}
Let $(\alg,\hil,D)$ be a {regular spectral triple} with simple dimension spectrum 
$\Sigma$, $E$ be a finite projective $\cA$-module
with Hermitian structure and Hermitian connection $\nabla$. 
In addtion, we assume that $0\notin \spec(D)$ and $0\notin\spec(D_{E, \nabla})$.
Then 
$(\End_\cA(E),E\otimes_\cA~\hil,$ $D_{E, \nabla})$ is also a regular spectral triple
with simple dimension spectrum $\Sigma'$ contained in 
$\Sigma -{\mathbb N}$. Moreover if $0 \notin \Sigma$, then $0 \notin \Sigma'$.
\end{proposition}

 To prove regularity we will use Higson's  characterization of regularity. For that we recall,

\begin{definition} Let $(\cA,\cH,D)$ be a spectral triple such that $\cA$ maps $\cH^\infty$ to itself. The {\em algebra of differential operators} $\cD(\cA,\Delta)$ is the smallest algebra of operators on $\cH^\infty$ closed under the operation $T \mapsto [\Delta,T]$ containing $\cA$ and $[D,\cA]$. Here $\Delta$ denotes $D^2$. 
\end{definition}
This algebra is filtered as follows, the elements of $\cA$ and $[D,\cA]$ have order zero and the operation $[\Delta, \cdot]$ raises order at most by one. Thus $\cD_k$, the space of operators of order at most $k$ are defined inductively as follows
\begin{eqnarray*}
\cD_0&=& \mbox{ algebra generated by } \cA {\mbox{ and }} [D,\cA] \\
\cD_1&=&\cD_0+[\Delta,\cD_0]+\cD_0[\Delta,\cD_0]\\
\cD_k&=&\cD_0+[\Delta,\cD_{k-1}]+\cD_0[\Delta,\cD_{k-1}]+\sum_{j=1}^{k-1} \cD_j \cD_{k-j}\\
\end{eqnarray*}
\begin{definition}
A spectral triple $(\cA,\cH,D)$ satisfies the {\em basic estimate} if  every differential operator $X \in \cD_k$ there is an $\epsilon >0$ such that
\[
\|D^kv\|+\|v\| \ge \epsilon \|X v\| 
\]
for all $v \in \cH^\infty$.
\end{definition}
\begin{theorem}[Higson, Theorem 4.26, \cite{Hi00}]
Let $(\cA,\cH,D)$ be a spectral triple such that every $a \in \cA$ maps $\cH^\infty$ to itself. Then this spectral triple is regular iff it satisfies the basic estimate. 
\end{theorem}

\begin{proof}
 For notational convenience let us denote $D_{E,\nabla}$ by $D'$. Let $\Delta'=D'^2$ and $ \cD(\End_\cA(E),\Delta') $ be the associated differential algebra with the filtration $\cD'$. Then note that ${\cD}^{\prime}_k \subseteq 
\cD_k$. 
Therefore $(\End_\cA(E),E\otimes_\cA \hil,D')$ satisfies the basic estimate hence regularity.

Now we proceed to the stability of the dimension spectrum. This is the final part of Proposition \ref{prop:enhancedRST}. We will divide this in two  cases. First let us tackle the case of free modules. Then if necessary by considering matrices with entries from $\cA$  we can assume $E=\cA$. 

\begin{lemma} \label{lemma:remainder} Let $T \in \cB$ and $0<\Re(\lambda)<\min\{d \in \sigma(\Delta)\} $, then\\
(a) $${(\lambda-\Delta)}^{-1}T= \sum_{k=1}^n \sum_{j=0}^k T_{jk}|D|^j (\lambda-\Delta)^{-1-k} +R_n(\lambda,T)$$
where $$T_{jk}= \binom{k}{j} 2^j \delta^{2k-j}(T) .$$\\
(b) $\|R_n(\lambda,T)\|_1 <C \max(|\Im{\lambda}|,1)^{-n/2}.$\\
(c) The function $z \mapsto \displaystyle \int_{C - i \infty}^{C+i \infty} \lambda^{-z} R_n(\lambda,T) D{(\lambda - \Delta)} ^{-1}d \lambda$ is holomorphic for large enough $n$ where $C$ is a real number separating zero from the spectrum of $\Delta$.
\end{lemma}

\begin{proof} (a) To prove (a) note,
\begin{eqnarray*}
(\lambda- \Delta)^{-1} T & = & T(\lambda-\Delta)^{-1}+ [(\lambda-\Delta)^{-1},T]\\
& = & T(\lambda-\Delta)^{-1}+ (\lambda-\Delta)^{-1}[\Delta,T](\lambda-\Delta)^{-1}\\
& = & \sum_{0 \le k \le n}T^{(k)}(\lambda-\Delta)^{-1-k}+(\lambda-\Delta)^{-1}T^{(n+1)}(\lambda-\Delta)^{-n}\\
\end{eqnarray*}
Here the last equality is obtained by iterating the previous one k-times and $T^{(k)}$'s are defined inductively as $T^{(0)}=T$ and $T^{(k)}=[\Delta,T^{(k-1)}]$.
For $b \in \cB$ we also have the following  relations:
\begin{eqnarray*}
[\Delta, b] & = & |D| \delta(b) + \delta(b) |D| \\
|D| b & = & b | D|  + \delta (b)\\
\end{eqnarray*}
Combining these two we get for $T \in \cB$, 
\begin{equation}
T^{(k)}=2 \delta(T^{(k-1)}) |D| + \delta^2(T^{(k-1)})\end{equation}
Applying this repeatedly we get 
\begin{align} \label{B:expn}
T^{(k)}&= \sum_{0 \le j \le k} \binom{k}{j} 2^j \delta^{2k-j}(T) |D|^j \\
& =  T_{jk}|D|^j.
\end{align}

(b)  To prove (b) note,
\begin{eqnarray*}
\||D|(\lambda-\Delta)^{-1/2}\| & \le & \sup_{d \in \sigma(|D|)} \frac{d}{|\lambda-d^2|}\\
& = &  \sup_{d \in \sigma(|D|)} {\sqrt{\frac{d^2}{(\Re(\lambda)-d^2)^2 +\Im(\lambda)^2}}}\\
& \le & \sup_{d \in \sigma(|D|)} {\sqrt{\frac{d^2}{(\Im(\lambda)^2(\Re(\lambda)-d^2)^2)}}} \qquad \text{if} \, 
\Im(\lambda)>1;\\
& \le & \frac{1}{{|\Re(\lambda)|^{1/4}}|(1-\sqrt{\Re(\lambda)|})}
\end{eqnarray*}
and is less than or equal to a constant when $\Im(\lambda)\le 1$.
Also, the trace norm $$||(\lambda - \Delta)^{-\alpha}||_1 \le \frac{C}{\max{(|\Im(\lambda) |, 1)}^{\alpha/2}}.$$
Therefore $ \||D|(\lambda-\Delta)^{-1/2}\| \le \frac{C}{\max{(|\Im(\lambda) |, 1)}}.$
Therefore one gets the trace norm estimate for $j\le n$,
\begin{align*}
\||D|^j(\lambda-\Delta)^{-n}\|_1 &\le \||D|^j(\lambda-\Delta)^{-j/2}\|\|(\lambda-\Delta)^{-(n-j/2)}\|_1  \\
& \le \frac{C_1}{\max{(|\Im(\lambda) |, 1)^{n/2}}}.\\
\end{align*}
Therefore the trace norm estimate for the remainder follows. \\
(c) Proof of (c) immediately follows from part (b).
\end{proof}

Let $D'=D+B$ for some $B \in \cB$, then $\Delta' = (D+B)^2 = D^2 +R = \Delta +R$ where $R=DB+BD+B^2$. Then by 
the resolvent identity, one has 
\begin{align}
(\lambda - \Delta')^{-1}  & = (1-(\lambda - \Delta)^{-1}R)^{-1} (\lambda - \Delta)^{-1} \\
& = (1-X)^{-1}(\lambda - \Delta)^{-1} \\
& = \sum_{0 \le k\le n} X^k (\lambda - \Delta)^{-1}  + X^{n+1}  (1 - X)^{-1}  (\lambda - \Delta)^{-1}  \label{resolvent}
\end{align}
where $X =  (\lambda - \Delta)^{-1} R$ and we have used the identity
$$ (1-X)^{-1} =  \sum_{k\le n} X^k + X^{n+1}  (1 - X)^{-1}.$$

By hypothesis, for $b \in \cB$, the function
$$
\Tr b |D|^{-2s} : s \to \Tr\left(b \int_{C-i\infty}^{C+i\infty} \lambda^{-s} (\lambda - \Delta)^{-1} d\lambda\right)
$$
is well defined on the half plane $\Re(s)>p$, and is analytic there, and has a 
meromorphic continuation to $\bbC$ with poles contained in $\Sigma$. Here
$C$ is a real number separating zero from the spectrum of $\Delta$ and $\Delta'$. 
We want to prove the same is true when $\Delta$ is replaced by $\Delta'$ and that is achieved by analyzing the meromorphic continuation of the difference $ \Tr b|D'|^{-2z}- \Tr b|D|^{-2z}$. For that note 
\begin{eqnarray*}
&&\Tr b|D'|^{-2z}- \Tr b|D|^{-2z}  \\
& = & \Tr b \left( \int_{C-i\infty}^{C+i\infty} \lambda^{-z} ((\lambda - \Delta')^{-1} -(\lambda - \Delta)^{-1} d\lambda \right)\\
&=& \Tr b \left( \int_{C-i\infty}^{C+i\infty} \lambda^{-z}(\sum_{1 \le k\le n} X^k (\lambda - \Delta)^{-1}  + X^{n+1}  (1 - X)^{-1}  (\lambda - \Delta)^{-1} )d \lambda \right) \\
& & (\mbox{ from \ref{resolvent}} )
\end{eqnarray*}

\begin{lemma} On any given right half plane, for  large enough $n$ the function 
$\, z \mapsto \Tr \left( b \displaystyle \int_{C-i\infty}^{C+i\infty} \lambda^{-z} X^{n+1}  (1 - X)^{-1}  (\lambda - \Delta)^{-1} d \lambda\right)$ is holomorphic.
\end{lemma}
\begin{proof} For a compact operator $T$ let $\mu_n(T)$ be the nth largest singular value of $T$. Let $d_n=\mu_n(D^{-1})$, then we have the following bounds
\begin{eqnarray*}
\mu_n(X) & < & C d_n\\
\|X\|=\mu_1(X) & < & C (|y|+1)^{-1}, \mbox{ where } y=Im(\lambda)\\
\end{eqnarray*}
We know $\sum d_n^{p+1}$ is finite. Therefore by Holder's inequality we get for large enough $n$ the trace norm of $X^{n+1}  (1 - X)^{-1}  (\lambda - \Delta)^{-1}$ is bounded by $(|\Im(\lambda)|+1)^{-k}$. From this the result follows. 
\end{proof}
\begin{lemma} For every $k \ge 1$,
$$ z \mapsto \Tr b \left( \int_{C-i\infty}^{C+i\infty} \lambda^{-z} X^k (\lambda - \Delta)^{-1} d \lambda \right)$$
defines a meromorphic function on $\bbC$ with poles contained in $\Sigma-{\mathbb N}:=\{ s-n: s \in \Sigma, n \ge 0\}$. Furthermore, this function is regular at zero.
\end{lemma}
\begin{proof} We will prove it for $k=1$, in the general case the proof is similar. In this case,
\begin{eqnarray*}
&& \Tr b \left( \int_{C-i\infty}^{C+i\infty} \lambda^{-z} X (\lambda - \Delta)^{-1} d \lambda \right)\\
&=& \Tr b \left( \int_{C-i\infty}^{C+i\infty} \lambda^{-z} (\lambda - \Delta)^{-1}DB (\lambda - \Delta)^{-1} d \lambda \right) \\ &+&  
\Tr b \left( \int_{C-i\infty}^{C+i\infty} \lambda^{-z} (\lambda - \Delta)^{-1}BD (\lambda - \Delta)^{-1} d \lambda \right) \\
&+& \Tr b \left( \int_{C-i\infty}^{C+i\infty} \lambda^{-z}(\lambda - \Delta)^{-1}B^2 (\lambda - \Delta)^{-1} d \lambda \right)\\
\end{eqnarray*}
We want to show that each term admits a meromorphic continuation to the complex plane with poles in $\Sigma-{\mathbb N}$. The first two terms are similar and we will only tackle the second term. From the analysis for the second term the result for the third term will follow. By Lemma \ref{lemma:remainder}, 

\begin{eqnarray*}
&&\Tr b \left( \int_{C-i\infty}^{C+i\infty} \lambda^{-z} (\lambda - \Delta)^{-1}BD (\lambda - \Delta)^{-1} d \lambda \right)\\
&= & \sum_{0 \le k \le n} \sum_{0 \le j \le k} \Tr b \left( \int_{C-i\infty}^{C+i\infty} \lambda^{-z} B_{jk}{|D|}^jD (\lambda - \Delta)^{-2-k} d \lambda \right)\\
&&+
 \Tr b \left( \int_{C-i\infty}^{C+i\infty} \lambda^{-z} R_n(\lambda, B) D (\lambda - \Delta)^{-1} d \lambda \right)\\
 &=& \sum_{0 \le k \le n} \Tr b \sum_{0 \le j \le k} B_{jk} \Sign(D)   \binom{-z}{k+1} {|D|}^{-z-k+j}\\
 && + {\mbox { a holomorphic function on a right half plane}}\\ 
 & = &  \sum_{0 \le k \le n}  \binom{-z}{k+1}\sum_{0 \le j \le k} \zeta (z+k-j,  b) \sum_{0 \le j \le k} B_{jk} \Sign(D) \\
 && + {\mbox { a holomorphic function on a right half plane}}\\ 
\end{eqnarray*}
where the first equality follows from Lemma \ref{lemma:remainder} $(a)$.
Clearly poles of this function are contained in $\Sigma-{\mathbb N}$.  The poles of this function are simple provided the zeta functions involved have simple poles. Note that if we assume that $\zeta (z,b)$ has simple poles for all $b \in \cB$ then we also get the regularity at zero.
 \end{proof}
 
 These lemmas give an alternative proof of the following result  in \cite{CC06}.
 
 \begin{cor}\label{cor:stability}
 Let $(\alg,\hil,D)$ be a {regular spectral triple} with simple dimension spectrum $\Sigma$.
 Suppose that $0 \notin {\rm spec}(D)$ and $0\notin {\rm spec}(D+T)$ for $T \in \cB_0$.
Then $(\alg,\hil,D+T)$ is also a regular spectral triple with simple dimension spectrum $\Sigma'$ contained in 
$\Sigma -{\mathbb N}$. Moreover if $0 \notin \Sigma$, then $0 \notin \Sigma'$.
 \end{cor}
 

 This proves Proposition \ref{prop:enhancedRST} for the case when $E$ is a finite free module. 
 Let $E = p\cA$, where $p$ is a projection in $\cA$, and $q = (1-p)$. Then $T = D - pDp - qDq \in \cB_0$ and $D+T = pDp + qDq$.
By the corollary above, we know that $(\alg,\hil,pDp + qDq)$ is also a regular spectral triple with simple dimension spectrum 
$\Sigma'$ contained in 
$\Sigma -{\mathbb N}$. Now the zeta function 
$$
\zeta(z, b) = \Tr(pbp|pDp + qDq|^{-z}) = \Tr(pbp|pDp|^{-z}), 
$$
 since $pq=0$. Therefore by Corollary \ref{cor:stability},
 $(p\cA p, p\cH, pDp)$ is also a regular spectral triple with simple dimension spectrum 
 $\Sigma'$ contained in 
$\Sigma -{\mathbb N}$, where we observe that $\End_\cA(E) = p\cA p$. This completes the proof of 
Proposition \ref{prop:enhancedRST}.
\end{proof}

\subsection{Removing the hypothesis, ``zero not in the spectrum''}\label{sect:zero}
We will now remove the hypotheses, $0 \notin {\rm spec}(D)$ and $0\notin {\rm spec}(D+T)$ in the 
 Corollary \ref{cor:stability} above. The difficulty lies in the fact that the orthogonal projection onto
 the nullspace of $D+T$, namely $P_0(D+T)$, is not in any obvious way in the algebra $\cB_0$, whenever
 $T\in \cB_0$. 
 So we will suitably enlarge the algebra $\cB_0$ to accommodate these projections, in such a way
 that the enlarged algebra does not alter the meromorphic continuation properties
 of the zeta functions.

Define the space of all smoothing operators 
$$\Psi^{-\infty} := \{ a \in B(\cH):  |D|^{k}a|D|^l  \quad\text{is bounded for all} \quad k, l \ge 0 \}. $$
Then $\Psi^{-\infty}$ is an ideal in the algebra $\{a \in B(\cH): a(\cH^\infty) \subset \cH^\infty \text{ and } a \in \cap_{n \ge 1} {\mathfrak D}om \delta^n\}$.

\begin{definition}
Let $\cB$ denote the algebra generated by $\cB_0$ and $\Psi^{-\infty} $.

\end{definition}

\begin{proposition}\label{prop:zeta} The zeta functions
$z\mapsto \zeta(z, b)= \Tr(b |D+P_0(D)|^{-z}), b \in \cB$ have the same meromorphic continuation properties as
that of $z\mapsto \zeta(z, b), b \in \cB_0$.
\end{proposition}

\begin{proof}
We begin by proving that for $b \in \Psi^{-\infty}$, $z\mapsto \zeta(z, b)$ is an entire function. 
This follows from $ \zeta(z, b) = \Tr(b|D+P_0(D)|^k |D+P_0(D)|^{-(k+z)})$ is holomorphic for $\Re(z)> p-k$, for all
$k\ge 0$, where we have also used the fact that $b|D+P_0(D)|^k$ is bounded whenever $b|D|^k$ is bounded,
for all $k \ge 0$. 

Notice that $ \Psi^{-\infty}$ is an ideal in $\cB$, therefore the proposition follows.
\end{proof}

\begin{lemma}\label{lemma:proj1}
The projections onto the nullspace of the operators of the form $D+T$ are smoothing operators, for all $T\in \cB_0$.
\end{lemma}

\begin{proof}
Since $P_0(D)$ is smoothing and $D^mP_0(D) = 0$.  
To show that $ P_0(D+T) $ is smoothing it is enough to show
that $D^m \left(P_0(D+T) - P_0(D)\right) D^l$ is bounded for all $ m, l \ge 0$.
Let $D'=D+T$ for some $T \in \cB$, then $\Delta' = (D+T)^2 = D^2 +R = \Delta +R$ where $R=DT+TD+T^2$.
Note that by the Cauchy formula we have
\begin{align*}
& D^m (P_0(D')-P_0(D)) D^l\\
& = \oint_{\cC_r}  d\lambda \,D^m\left( (\lambda - \Delta')^{-1} -  (\lambda - \Delta)^{-1} \right) D^l\\
& =  \oint_{\cC_r}  d\lambda \,D^m\left( \sum_{1 \le k\le n} X^k (\lambda - \Delta)^{-1}  + X^{n+1}  (1 - X)^{-1}  (\lambda - \Delta)^{-1} \right) D^l
\end{align*}
where $X =  (\lambda - \Delta)^{-1} R$ and we have used the identity
$$ (1-X)^{-1} =  \sum_{k\le n} X^k + X^{n+1}  (1 - X)^{-1}.$$
Here $\cC_r$ is a circle of radius $r$ centered at the origin in $\bbC$, where $r>$ is small.

We analyse  the expression when $k=1$, as well as the remainder term, as the other terms 
can be analysed similarly.
$$\oint_{\cC_r}  d\lambda \,D^m   (\lambda - \Delta)^{-1} (DT+TD+T^2) (\lambda - \Delta)^{-1} D^l$$
is equal to 
$$\oint_{\cC_r}  d\lambda \,D^m  (\lambda - \Delta)^{-1}  ([D,T]+2TD+T^2) (\lambda - \Delta)^{-1} D^l$$
We analyse the $TD$-term. 
$$\oint_{\cC_r}  d\lambda \,D^m  (\lambda - \Delta)^{-1}  TD (\lambda - \Delta)^{-1} D^l$$
Applying part (a) of Lemma \ref{lemma:remainder}, we obtain
 \begin{align*}
& \sum_{0 \le k \le n} \sum_{0 \le j \le k}  \oint_{\cC_r} D^m T_{jk}{|D|}^jD (\lambda - \Delta)^{-2-k} D^ld \lambda\\
& +  \oint_{\cC_r} D^m R_n(\lambda, B) D (\lambda - \Delta)^{-1} D^l d \lambda.
\end{align*}
 Upon commuting $D^m$ and $T_{jk}$, at the expense of adding more commutator terms,
we get  terms of the form,
$$T_{jk} \oint_{\cC_r} D^{m+l+j+1} {\rm sign}(D)  (\lambda - \Delta)^{-2-k} d \lambda$$
which clearly vanish, together with a remainder term,
$$\oint_{\cC_r} D^m R_n(\lambda, B) D (\lambda - \Delta)^{-1} D^l d \lambda.
$$
The remainder is bounded since it contains $  (\lambda - \Delta)^{-1-n}$ which dominates 
$D^m$, whenever $m+l\le n+1$.
\end{proof}

\begin{lemma}\label{lemma:proj2}
The projections onto the nullspace of the operators $D_{E, \nabla}$ are smoothing operators.
\end{lemma}

\begin{proof}
 Let $E = p\cA$, where $p$ is a projection in $\cA$, and $\nabla_0=pdp$ be the 
 Grassmanian connection. We need to first show that  the projection onto the nullspace of $D_{E, \nabla_0} = pDp$ is
 a smoothing operator. If $q = (1-p)$, then $T= pDp+qDq-D \in \cB_0$. It follows that 
 $pDp+qDq$ is an operator of the form as in Lemma \ref{lemma:proj1}. Therefore 
the projection onto the nullspace of  $pDp+qDq$ is a smoothing operator. But $P_0(pDp)$
is just $pP_0(D)$, which is smoothing since $p \in \cB_0$ and smoothing operators form an ideal. 
 A general connection $\nabla$ on $E$ is of the form $\nabla_0 + T$ and so $D_{E, \nabla}$
 is of the form $D_{E, \nabla_0} + T$, and applying Lemma \ref{lemma:proj1}, we deduce the 
 Lemma.
\end{proof}

Proposition \ref{prop:zeta} and Lemmas \ref{lemma:proj1} and \ref{lemma:proj2} prove Proposition \ref{prop:enhancedRST2}, which is just Proposition \ref{prop:enhancedRST}
with the hypotheses ``zero not in the spectrum'' removed.

\section{An example}

Here we study the case of the noncommutative torus,
for which the family of spectral triples is regular, and therefore the determinant line bundle $\cL$ has a 
Quillen metric and determinant section over the 
moduli space of flat rank 1 connections, which we calculate explicitly in terms of theta and eta functions on the 
moduli space which is a torus.

Recall that the {\em noncommutative torus} $A_\theta$ is the universal C$^*$-algebra 
generated by two unitaries $U, V$ satisfying the Weyl commutation relations,
$UV = e^{i\theta} VU$ for fixed $\theta \in \bbR$. There is a 
natural smooth subalgebra
$A^\infty_\theta$ called the {\em  smooth noncommutative torus}, 
which is defined as those 
elements in $A_\theta$ that can be represented by infinite
power series 
$
f = \sum_{(m, n)\in \ZZ^2} a_{(n, m)} \, U^m V^n,
$
with $(a_{(m, n)})\in 
\cS(\ZZ^2)$, the Schwartz space of rapidly decreasing 
sequences on $\ZZ^2$.

We next briefly recall the space of differential forms on $A^\infty_\theta$
which is defined by $\Omega^j(A^\infty_\theta) = 
A^\infty_\theta \otimes \Lambda^j(\bbC^2)$. We remark that there are two notions of differential forms (\cite{Connes94}, \cite{CR}) and in the context of the noncommutative torus one knows \cite{Fro99} through explicit computations  that they agree.
The differential $d$ is defined as 
$d(a \otimes \alpha) := \delta_1 (a)  \otimes e_1\wedge\alpha +  \delta_2 (a)  \otimes e_2\wedge\alpha
$
where $a \in A^\infty_\theta, \,  \alpha \in \Lambda^\bullet(\bbC^2),$ and $e_j , j = 1, 2$ is the canonical basis of 
$\bbC^2$.  Here the standard derivations are defined on the generators by
$ \delta_1 (U)= 2\pi U, \,  \delta_2 (V)= 2\pi V, \,  \delta_1 (V) =  \delta_2 (U) = 0$.

We now consider the space of all flat Hermitian connections on the trivial rank one free module
over $A^\infty_\theta$. The space of all such connections is an affine space $\cC_{A^\infty_\theta}$ with
associated vector space $Z^1(A^\infty_\theta)$, which is the vector space of closed 1-forms on $A^\infty_\theta$.
Let $U(A^\infty_\theta)$ denote the space of all unitary elements in $A^\infty_\theta$. Then 
by Proposition 5.7 in \cite{CR}, we know that the quotient space  $\cC_{A^\infty_\theta}/U(A^\infty_\theta)$ 
is homeomorphic to the torus ${\mathbb T}^2$. This is the analogue of the {\em Jacobian variety}, for the 
noncommutative torus $A^\infty_\theta$.

We next recall the spectral triple on $A^\infty_\theta$, cf \cite{Varilly}, Section 4. 
Let $\cH = \cH^+ \oplus \cH^-$, where $\cH^\pm$ are both copies of $L^2(A_\theta, \tr)$,
where $\tr$ denotes the canonical trace on $A_\theta$.
Fix $\tau\in \bbC$ such that $\Im(\tau)>0$, 
and define $\partial = \partial_\tau = \delta_1 + \tau \delta_2$, so that 
$\partial^* = -\delta_1 - \bar\tau \delta_2$. Finally, let $D = 
\left(\begin{array}{cc}0 &  \partial^*\\\partial & 0\end{array}\right)$. Then 
a spectral triple for $A^\infty_\theta$ is $(A^\infty_\theta, \cH, D)$. From what follows,  
this is a {\em regular spectral triple}.

The flat Hermitian connections are $\nabla = \nabla_\tau = (\delta_1 + u)\otimes e_1 + \tau (\delta_2 +v)\otimes e_2$ 
where $u, v \in [0, 1)$, being thought as parametrizing the Jacobian variety. 
Then setting $\partial_\nabla =  (\delta_1 + u) + \tau (\delta_2 + v)$, we see that 
$\partial_\nabla^* = -  (\delta_1 + u) - \bar\tau (\delta_2 + v)$ and  $D_\nabla = 
\left(\begin{array}{cc}0 &  \partial_\nabla^* \\ \partial_\nabla & 0\end{array}\right)$. Then 
we notice that $D^2 = \left(\begin{array}{cc} \partial_\nabla^*\partial_\nabla &  0 \\ 
0 & \partial_\nabla\partial_\nabla^* \end{array}\right)$ and the vectors $U^mV^n$
form an orthonormal basis of eigenvectors for both $\partial_\nabla^*\partial_\nabla$ and 
$ \partial_\nabla\partial_\nabla^* $, and we compute  the eigenvalues,
\begin{align*}
\partial_\nabla^*\partial_\nabla(U^mV^n) & =\partial_\nabla\partial_\nabla^*(U^mV^n) \\
& = - ( (\delta_1 + u) + \tau (\delta_2 +v)) ((\delta_1 + u) + \bar \tau (\delta_2 +v))(U^mV^n) \\
& = - \frac{4 \pi^2}{\Im(\tau)^2} |(m + u) - \tau (n + v)|^2.
\end{align*}
The associated zeta function 
$$
\zeta(s) = \sum_{m, n} \frac{\Im(\tau)^{2s}}{4^s \pi^{2s}} |(m + u) - \tau (n + v)|^{-2s}.
$$
is holomorphic for $\Re(s)>>0$, and is precisely the $\zeta$-function considered 
in \cite{RS} in the proof of Theorem 4.1 in that paper, where it arose in a completely
different context. Using the results from there, 
it follows that $\zeta(s)$ has a
meromorphic continuation to $\bbC$, with no pole at $s=0$, showing in 
particular that the spectral triple that we started out with on the noncommutative 
torus is a {\em regular spectral triple} such that zero is not in the dimension spectrum. 
Moreover, the Quillen norm of the determinant section of the determinant line bundle $\cL$ over the Jacobian variety is 
$$
{\mathbb T}^2 \ni (u, v) \mapsto ||\det(D)||_\cL (u, v) = e^{-\zeta'(0)} =\left| e^{\pi i v^2 \tau}\frac{ \theta_1(u-\tau v, \tau) }{\eta(\tau)}\right| \in {\mathbb R}^{\ge 0},
$$
where the theta function is defined as 
$$
\theta_1(w, \tau) = -\eta(\tau) e^{\pi i (w + \tau/6)} \prod_{k=-\infty}^\infty (1- e^{2\pi i (|k|
\tau - \epsilon_k w)}, 
$$
where $\epsilon_k = {\rm sign}\left(k+\frac{1}{2}\right)$, and the Dedekind eta function is 
defined by 
$$
\eta(\tau) = e^{\pi i \tau/12} \prod_{k=1}^\infty (1- e^{2\pi i k \tau}), 
$$
From these formulae, one immediately recovers the determinant section as
in the text of the paper.  Also by  the explicit knowledge of the eigenvalues 
and the determinant, 
$$
e^{-\zeta'_\lambda(0)} = 
\left| e^{\pi i v^2 \tau}\frac{ \theta_1(u-\tau v, \tau) }{\eta(\tau)}\right|  \left(\prod_{0<  q <\lambda}  \frac{4 \pi^2}{\Im(\tau)^2} |(m + u) - \tau (n + v)|^2\right)^{-1}
$$
where $q =  \frac{4 \pi^2}{\Im(\tau)^2} |(m + u) - \tau (n + v)|^2$, determining the Quillen 
metric on $\cL$ in this case over the open subsets $U_\lambda$ of the Jacobian torus.\\

{\bf Acknowledgements:} Both authors gratefully acknowledge support from the Australian Research Council.

\end{document}